\documentstyle[12pt,psfig]{article}
\input{epsf}
\begin{document}
\hfill UFTP-498/1999
\vspace*{0.3 truecm}
\begin{center}
{\LARGE \bf Studying Phase Transitions in Nuclear Collisions}\\
\vspace{0.5 truecm}
{\Large \bf I.N. Mishustin}\\
\vspace{0.5 truecm}
{\small \it The Kurchatov Institute, Russian Research Center, 123182 Moscow,
Russia;}\\  
{\small \it The Niels Bohr Institute, Blegdamsvej 17, DK-2100 Copenhagen {\O},
Denmark;}\\
{\small \it Institute for Theoretical Physics, J.-W. Goethe University, 
Robert-Meyer Str. 8-10, D-60054 Frankfurt am Main, Germany}
\end{center}
%\vspace{0.3 truecm}
\begin{abstract}
{\small
In this talk I discuss three main topics concerning the theoretical description 
and observable signatures of possible phase transitions in nuclear collisions.
The first one is related to the multifragmentation of equilibrated sources and 
its connection to a liquid-gas phase transition in finite systems. The second
one is dealing with the Coulomb excitation of ultrarelativistic
heavy ions resulting in their deep disintegration. The third topic is devoted to
the description of a first order phase transition in rapidly expanding matter. 
The resulting picture is that a strong collective flow of matter will lead to
the fragmentation of a metastable phase into droplets. If the transition from
quark-gluon plasma to hadron gas is of the first order, it will manifest itself
by strong nonstatistical fluctuations in observable hadron distributions.}
\end{abstract}
\vspace{0.3 truecm}

\begin{center}
{\large \bf INTRODUCTION} 
\end{center}

A general goal of present and future experiments with heavy-ion beams is to study
the properties of strongly interacting matter away from the nuclear ground
state. The main interest is focussed on searching for and studying possible 
phase transitions. Several phase transitions are predicted in different domains 
of temperature $T$ and baryon density $\rho_B$. There is no doubt that 
there should be a first order phase transition of the liquid-gas type in 
normal nuclear matter. This follows simply from the existence of the nuclear 
bound state at the saturation density $\rho_0 \approx 0.15$ fm$^{-3}$.
Therefore, at $\rho_B<\rho_0$ and low temperatures, $T<T_c \sim 10$ MeV, the matter 
will organize itself in the form of a mixed phase with droplets of nuclear
liquid surrounded by the nucleon gas. The only problem is whether relatively
small amounts of excited nuclear matter produced in nuclear collisions and its 
limited lifetime are sufficient to observe this phase transition. Based on 
recent data on the nuclear caloric curve \cite{bondorf3} and temperature 
fluctuations \cite{Dago99} I am 
tempting to give a positive answer to this question. This topic will be
discussed in the first part of the talk after a short description of the
Statistical Multifragmentation Model (SMM) \cite{I,Mish85} which provides a 
basis for theoretical analysis.

The situation at high $T$ and nonzero baryon chemical potential $\mu$ 
($\rho_B>0$) is not so clear, although everybody is
sure that the deconfinement and chiral transitions should occur somewhere.  
The phase structure of QCD is not yet fully understood. Reliable lattice
calculations exist only for $\mu=0$ ($\rho_B=0$) where they predict a 
second order phase transition or crossover at $T\approx 160$ MeV. 
As model calculations show, the phase diagram in the
$(T,\mu)$ plane may contain a first order transition line (below called 
the critical line) terminated at a (tri)critical point \cite{Raj,Jac}. 
Possible signatures of this point in heavy-ion collisions 
are discussed in ref. \cite{Ste}. Under certain non-equilibrium conditions, 
a first order transition is also predicted for symmetric quark-antiquark 
matter \cite{Sat}.

A striking feature of central heavy-ion collisions at high energies, confirmed
in many experiments (see e.g. \cite{reisdorf,Bra}), is a very strong collective
expansion of matter. The applicability of equilibrium concepts for
describing phase transitions under such conditions becomes questionable. 
In the last part of the talk I demonstrate that non-equilibrium phase 
transitions in rapidly expanding matter can lead to interesting 
phenomena which, in a certain sense, can be even easier to observe \cite{Mish0}.

In the middle part of the talk I address a question which is closely related to the
main topic of this conference. It illustrates how the knowledge accumulated in 
intermediate-energy heavy-ion physics can be used for ultrarelativistic 
heavy-ion colliders. Namely, I will discuss the excitation of nuclei by 
Lorentz-contracted and strongly-enhanced Coulomb fields
of ultrarelativistic heavy ions. As well known, this process can be treated in
terms of equivalent photons. Their flux grows linearly with the
squared nuclear charge, and their characteristic energy is proportional to 
the relative Lorentz factor of colliding nuclei. This is why this Coulomb
excitation of nuclei becomes especially important in high-energy heavy-ion
colliders such as RHIC and LHC. The calculations \cite{Pshe1,Pshe2} show that 
in such colliders the equivalent photon spectrum extends far above the Giant 
Resonance region, into the GeV domain. The absorption of such a photon by a 
nucleus leads to its high excitation and subsequent disintegration. This might 
be an important factor determining a lifetime of ultrarelativistic heavy-ion beams.
\\

\begin{center}
{\large \bf STATISTICAL MULTIFRAGMENTATION AND LIQUID-GAS PHASE TRANSITION}
\end{center}

When a nucleus is suddenly heated up to a temperature $T$ it starts expanding 
to adjust a new equilibrium density $\rho_0(T)$ which is less than the 
equilibrium density at zero temperature $\rho_0$. If the initial temperature
is high enough the expansion is unlimited. At some stage of expansion the 
system enters into the spinodal region, where 
the  homogeneous distribution of matter becomes  thermodynamically unstable. 
Therefore, the nucleons form smaller and bigger clusters or droplets with density close to 
$\rho_0$. This clusterization process resembles a liquid-gas phase transition 
in ordinary fluids. In the transition region the matter is very soft in a 
sense that the sound velocity is close to zero (soft point). This means also 
that the expansion is slow and the system has enough time to find a most
favorable cluster-size distribution maximizing the entropy. 

At a later stage of expansion the system reaches a so-called freeze-out
state when clusters cease to interact with each other. This break-up state of the 
system can be described within a statistical approach. In 1985 we have
constructed a Statistical Multifragmentation Model (SMM) \cite{I,Mish85} which 
up to now  is one of the most successful realizations of this approach for 
finite nuclear systems. The model and its numerous applications are described in 
detail in a recent review \cite{PR}. A similar model was also constructed by
Gross \cite{gross}.  In this talk I outline only some general 
features of the SMM and give a few examples of how it works. 

It is assumed that at break-up the system consists of primary hot 
fragments and nucleons in thermal equilibrium. Each break-up channel or
partition, $f$, is specified by the multiplicities of different species, 
$N_{AZ}$, constrained by the total baryon number $A_0$ and charge 
$Z_0$. 
The total fragment multiplicity is defined as $M=\sum_{AZ} N_{AZ}$.
The probabilities of different break-up channels are calculated 
in an approximate microcanonical way according to their statistical weights,
\begin{equation}
W_f \propto \exp\left[S_f(E^*,V,A_0,Z_0)\right], 
\end{equation}
where $S_f$ is the entropy
of a channel $f$ at excitation energy $E^*$ and volume $V$. 

Translational degrees of freedom of fragments are described by the Boltzmann 
statistics while the internal excitations of individual fragments with 
$A>4$ are calculated according to the quantum liquid-drop model. 
An ensemble of microscopic states corresponding to a break-up channel $f$ 
is characterized by a temperature $T_f$ which is 
determined from the energy balance equation
\begin{equation} \label{enbal}
\frac{3}{2}T(M-1)+\sum_{(A,Z)}E_{AZ}(T)N_{AZ}+E_f^C(V)-Q_f=E^*~.
\end{equation} 
Here the first term comes from the translational motion, 
the second term includes internal excitation energies of individual fragments, 
the third term is the Coulomb interaction energy and the last one is the Q-value
of the channel $f$. The excitation energy 
$E^*$ is measured with respect to the ground state of the compound nucleus 
($A_0$,$Z_0$). It is fixed for all fragmentation channels while the 
temperature $T_f$ fluctuates from channel to channel.
 
The total break-up volume is parametrized as $V=(1+\kappa) V_0$,
where $V_0$ is the compound nucleus volume at normal density and the 
model parameter $\kappa$ is the same for all channels. 
The entropy associated with the translational motion of fragments is 
determined by the  ``free'' volume, $V_f$, which is only a fraction of 
the total break-up volume $V$. In the SMM $V_f(M)$ is 
parametrized in such a way that it grows almost linearly with the primary 
fragment multiplicity $M$ or equivalently, with the excitation energy 
$\varepsilon^*=E^*/A_0$ of the system \cite{PR}. 

At given inputs $A_0$, $Z_0$ and $\varepsilon^*$ the individual multifragment 
configurations are generated by the Monte Carlo method. After the break-up 
the hot primary fragments propagate in a common Coulomb field and loose their 
excitation. The most important de-excitation mechanisms included in the SMM 
\cite{PR} are the simultaneous Fermi break-up of lighter fragments  ($A\leq 
16$) and the evaporation from heavier fragments, including the compound-like  
residues. In refs. \cite{botvina1,xi,dagostino,boug,viola,sri} one can find fresh
examples showing how well the SMM works in describing the multifragmentation 
of thermalized sources.

In ref. \cite{das} an equation of state of a multifragment system
was calculated for the grand canonical version of the SMM. As expected, it 
shows clear signs of a liquid-gas phase transition with a critical temperature 
of about 7 MeV. In the transition region the pressure isotherms are very flat
indicating that the sound velocity is very small. In this region the 
compressibility and specific heat have nonmonotonic behaviour.   

The most interesting prediction of the statistical model, a plateau in the 
caloric curve $T(\varepsilon^*)$, was formulated already in 1985 \cite{bondorf}.
Since that time it was a challenge for experimentalists to measure the 
nuclear caloric curve. First  measurements were performed 
at GSI by the ALADIN collaboration only in 1995 \cite{pochod}. They have 
shown an impressive agreement with the theoretical prediction. These 
results have initiated an  avalanche of other measurements and a lively 
discussion in the community (see latest ALADIN results in ref. \cite{mul}).  

Most temperature measurements are based on the 
Albergo method \cite{albergo} relating the temperature to the double ratio
of isotope yields. The analysis shows (see for instance \cite{bondorf3,tsang}) 
that the temperatures extracted by this method are very sensitive to
the side-feeding and nuclear structure effects. 
According to SMM the observed light isotopes are produced mainly by 
the secondary decays of hot primary fragments. This leads to a difference 
between the isotopic temperatures and true thermodynamical temperature at
freeze-out (see detailed analysis in ref. \cite{bondorf3}). In particular, isotopic
temperatures have typically a less pronounced plateau than the true temperature,
which can even have a backbending. In recent years several
comparisons have been made (see examples in refs. \cite{bondorf3,xi,raduta} 
which generally show very good agreement between the theory and experiment.

\begin{figure}
%\begin{center}
\vspace{-2.5cm}
\hspace{2cm}
\mbox{
\epsfysize=15cm
\epsffile{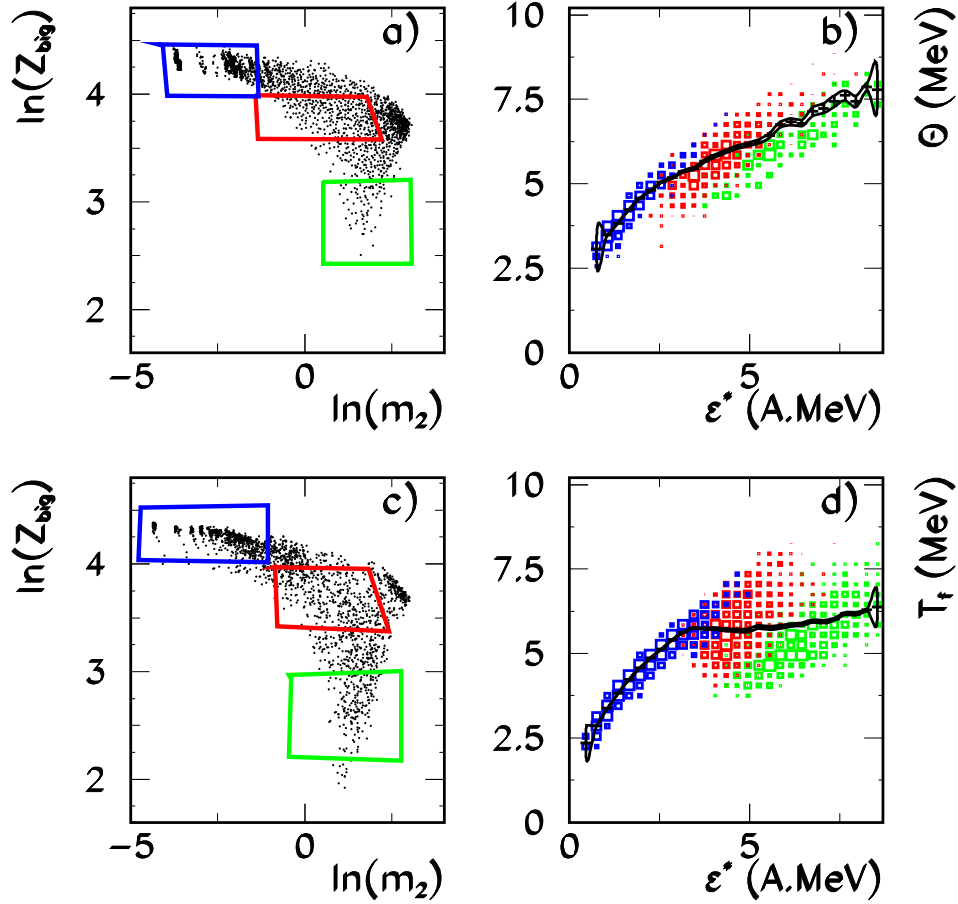}
}
\caption{ \small
Experimental (a) and SMM generated (c) Campi scatter plots  Three cuts
are introduced to select liquid-like events (Cut 1), gas-like events (Cut 3) and
critical events (Cut 2). Panels b) and d) show the correlation between 
temperatures and excitation energies for experimental and SMM generated events. 
Events belonging to different cuts are shown by blue (1), red (2) and green (3)
squares. Sizes of the squares are proportional to the yields. 
%The regions populated by the events from the three cuts are encircled. 
The solid lines in panels b) and d) show the mean temperatures for all the 
events at given $\varepsilon^*$.  }
%\end{center}
\end{figure}

One should bear in mind that the ALADIN caloric curves are measured for a wide 
ensemble of decaying sources associated with the projectile or target
spectators produced in peripheral nuclear collisions. The thermodynamical
significance of such observations would increase if the measurements were done
for a fixed source size with a varying excitation energy. Also the temperature
measurements on the event-by-event basis would make it possible to study its
fluctuations and therefore the heat capacity of the nuclear system. Such an 
analysis was performed recently by the Bologna group \cite{Dago99} in the study 
of quasi-projectile (QP) fragmentation in peripheral Au+Au collisions at 35 A MeV.  
In this analysis only the events with reconstructed QP charges $70<Z_{QP}<88$
were included. The excitation energy, determined by a calorimetric method,
varied for these events from 0.5 to about 8 Mev/nucleon.

Measuring temperatures event by event is of course a nontrivial task. 
An attempt of estimating the event ``temperature''
(below denoted by $\theta$) was made in ref. \cite{Dago99}.
The idea is to apply the energy 
balance equation (\ref{enbal}), which is used in the SMM, but now  for the
experimental events. Of course, this requires certain assumptions on how the 
observed partitions, involving cold reaction products, are related to the 
original partitions consisting of hot primary fragments.  Therefore, it was
assumed that the light particles detected in a partition were produced by
de-excitation of hot primary fragments. For reconstructing a primary partition
these light particles were shared among the detected fragments proportionally 
to their charges and assuming the charge-to-mass ratio as in the entrance
channel. Applying this procedure for the asymptotic SMM events showed that
the correlation between the microcanonical temperature and excitation energy 
was reproduced within 5$\%$. 
  
Fig.~1 shows the scatter plots in the $(T,\varepsilon^*)$ plane for experimental
(b) and SMM generated (d) events. The ensemble-averaged 
temperatures are indicated by the solid lines. Their behaviour is typical for
caloric curves measured by other methods. In addition to the flattening of the
average temperature at $T\approx $ 6 MeV,  one can clearly see the broadening 
of the distributions in the transition region at $\varepsilon^*=4\div 8$ 
MeV/nucleon. The quantity characterizing energy fluctuations is heat capacity. 
For a canonical ensemble at constant volume it can be expressed as 
\begin{equation} \label{cv}
C_V=\frac{\sigma_E^2}{T^2}=\frac{\langle E^2\rangle-\langle E\rangle^2}{T^2}.
\end{equation}
It is not clear whether the constant volume condition applies to actual
freeze-out configurations but nevertheless studying the energy fluctuations 
provides an additional and important information compared to the average
characteristics. Indeed, applying Eq. (\ref{cv}) for scatter plots of Fig.~1
reveals a peak in $C_V$ at temperatures around 6 MeV. This behaviour was 
also predicted theoretically a long time ago \cite{bondorf,gross}.

Another way of characterizing the critical behaviour is to analyze the conditional
moments of fragment multiplicity distributions introduced by Campi \cite{campi}. 
Fig. ~1 a) shows, for each event $j$, the experimental correlation between
the logarithm of the charge of the largest fragment, $\ln{(Z_{big}^{(j)})}$, 
and the logarithm of the corresponding second moment 
of the multiplicity distribution, $\ln{(m_2^{(j)})}$ (Campi scatter plot). 
Fig.~1 c) shows the same for events generated by
the SMM. As expected for a system experiencing a phase transition these plots 
exhibit two branches: an upper branch with an average negative slop,
corresponding to under-critical events, and a lower branch with a positive slop
that corresponds to super-critical events. The two branches meet in a central
region signalling the approach to a critical point. This trend is nicely
reproduced in Fig.~1 by both the experiment and the theory. 

We have made three cuts in these scatter plots selecting the upper branch (Cut 
1), the lower branch (Cut 3) and the central region (Cut 2) and analyzed the 
events falling in each of the three zones.
The fragment charge distributions in these three zones exhibit shapes going from
a U-shape in Cut 1, characteristic of the evaporation events at low excitation
energies, to an exponential one in Cut 3, characteristic of the vaporization
events at high excitations. In Cut 2 a power-low fragment charge distribution
Z$^{-\tau}$ with $\tau \approx 2.2$ is observed as expected according to the
Fisher's droplet model for fragment formation near the critical point of a
liquid-gas phase transition \cite{fisher} (see also an interesting analysis of
ref. \cite{sri}). 

The contributions of these three types of events to the caloric curves are shown 
in Fig.~1 for experiment (b) and for theory (d). It is clearly seen for both 
the data and the SMM, that in Cuts 1
and 3 besides normal events there are unusual events (although with low
probability) which lie far from the average $T(\varepsilon^*)$ behaviour.
These are compound-like states with very high temperatures and vaporization
events with low temperatures. For these events one can make the analogy  
respectively  with an overheated liquid and a super-cooled gas in the ordinary
liquid-gas phase transition. Here we see the advantage of a finite system where
not only the most probable states but also the metastable states can be produced
with a finite probability. In my opinion, the observation of these metastable 
states is the best indication that we are dealing here with the first order
phase transition of the liquid-gas type. These interesting questions were
further studied in ref. \cite{chomaz}.
\\

\begin{center}
{\large \bf ELECTROMAGNETIC EXCITATION OF ULTRARELATIVISTIC HEAVY IONS}
\end{center}

It has become clear in recent years \cite{Pshe1,Pshe2} that high nuclear 
excitations can be induced by the Coulomb fields of ultrarelativistic heavy
ions. Following the famous Waizs\"acker-Williams method the Lorentz contracted 
Coulomb field of an ultrarelativistic projectile in the rest frame of a target
nucleus (and vice versa) can be represented as a beam of equivalent or virtual
photons. The flux of equivalent photons with energy $E_\gamma$ in a collision 
of nuclei with charge $Z$ at impact parameter $b$ is given by the standard 
formula 
\begin{equation}
N(E_\gamma, b)=\frac{\alpha Z^2}{\pi^2}\frac{x^2}{\beta^2 E_\gamma b^2}
\left[K_1^2(x)+\frac{1}{\gamma^2}K_0^2(x)\right],
\end{equation}
where $\alpha$ is the fine structure constant, $\beta=v/c$ and 
$\gamma=\sqrt{1-\beta^2}$ is the 
relative Lorentz factor. The variable $x$ in the modified Bessel 
functions $K_{0,1}(x)$  is defined as 
$
x=E_\gamma b /(\beta \gamma \hbar c).
$
Since $K_{0,1}$ drop exponentially at large arguments, 
the main contribution to the virtual photon flux comes from the 
region $x\sim 1$. Thus the characteristic energy of virtual
photons grows linearly with $\gamma$. This explains why the relativistic
Coulomb excitation is very important for ultrarelativistic heavy-ion beams
where both $\gamma$ and $Z$ are large. For colliding beams
$\gamma=2\gamma_{beam}^2-1$ that gives $2\cdot 10^4$ and $10^7$ for RHIC and 
LHC respectively. This brings the spectrum of virtual photons
into the GeV energy domain, i. e.  much above the traditionally studied Giant 
Resonance (GR) and Delta-resonance regions.
The absorption of such a high-energy photon will result in a very high nuclear
excitation sufficient for its total disintegration.

In ref. \cite{Pshe0} the description of nuclear photoabsorption was
extended to the photon energies much above the GR region, where the excitation
of individual nucleons and multiple pion production are the dominant reaction
channels. A model of electromagnetic dissociation (ED) taking into account these 
high-energy photon absorption channels was constructed in ref. \cite{Pshe1}. 
According to this model, the fast hadrons produced after the photon absorption 
initiate a cascade of subsequent collisions with the intranuclear nucleons
leading to the fast particle emission and heating of a residual nucleus. This 
stage is described by the
Intranuclear Cascade Model (INC). At a later stage the nucleus
undergoes de-excitation by means of the evaporation of nucleons and lightest
fragments, binary fission or multifragmentation. The latter process becomes
important at ultrarelativistic beam energies, when the excitation energy of 
residual nuclei exceeds 3-4 MeV/nucleon. This stage of the reaction is
described by the SMM. 

\begin{figure}
\vspace{-1.2cm}
\begin{center}
\mbox{
\epsfysize=11cm
\epsffile{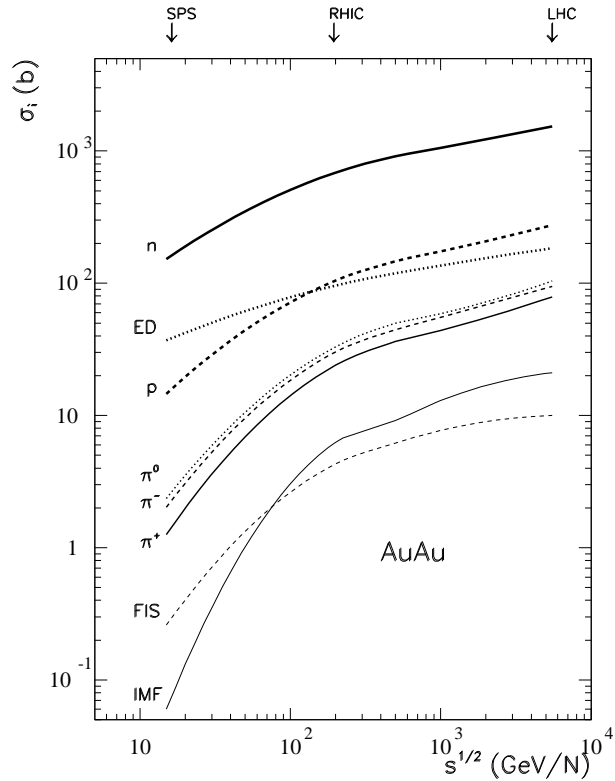}
}
\caption{ \small
Theoretical predictions for inclusive cross sections for emitting nucleons,
pions, intermediate mass fragments (IMF: $3\le Z \le 30$) and fission fragments
(FIS: $30< Z \le 50$) in the electromagnetic dissociation of Au nuclei as
functions of the c.m. energy $\sqrt{s}$ (the SPS, RHIC and LHC energies are
indicated by arrows). The thick dotted line shows the total ED cross
section for Au beams. } 
\end{center}
\end{figure}
\vspace{0.3cm}

To include all the processes described above, in ref. \cite{Pshe2} we have 
developed a specialized computer code RELDIS aimed at the Monte Carlo simulation
of the Relativistic ELectromagnetic DISsociation of nuclei. The simulation
begins with generating the single- or double-photon absorption process. Then the
INC model is used to calculate the fast particle emission and the
characteristics of residual nuclei. Finally, de-excitation of thermalized
residual nuclei is simulated by the SMM.  

The cross section of the photo-nuclear ($\gamma A$) reaction induced by a
photon of energy $E_\gamma$ is expressed as 
\begin{equation} \label{EDcs}
\frac{d\sigma_{ED}}{dE_\gamma}=\sigma_{\gamma A}(E_\gamma)\int_{b_{min}}^\infty
N(E_\gamma,b)2\pi bdb,
\end{equation}
where $b_{min}\approx (R_p+R_t)$ is a minimal impact parameter for heavy-ion 
collisions without nuclear overlap, $\sigma_{\gamma A}(E_{\gamma})$ is an appropriate
photo-absorption cross section, either measured for the $A$-nucleus with real
photons or calculated within a model. The total ED cross section,
$\sigma_{ED}$, is obtained
by integrating Eq. (\ref{EDcs}) by $dE_\gamma$ from 0 to $\infty$.
The calculations \cite{Pshe1} show that the total ED cross
sections for RHIC and LHC are very large, 100 b and 200 b respectively.
Accordingly, the ED reaction rates are much higher than those for nuclear
interactions, although the ED events are much less violent. 
For instance, at expected RHIC luminosity  $L\approx
10^{27}$ cm$^{-2}$s$^{-1}$ the ED reaction rate will be $10^5$ interactions per
second. Together with the electron capture reactions the ED processes will be
the important factors reducing the lifetime of ultrarelativistic heavy-ion beams
compared with the proton ones.  

\begin{figure}
\begin{center}
\vspace{-2cm}
\mbox{
\epsfysize=10cm
\epsffile{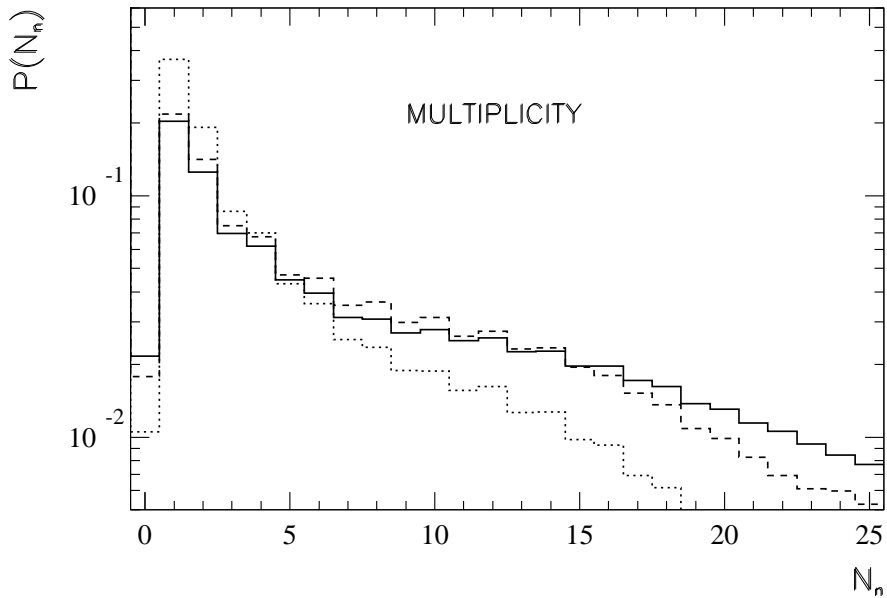}
}
\caption{\small
Normalized neutron multiplicity distributions for the 
electromagnetic
dissociation of Pb nuclei at LHC and SPS (solid and dotted histograms,
respectively) and Au nuclei at RHIC energies (dashed histogram).
Calculations are made with RELDIS code.}
\end{center}
\end{figure}

We have applied the RELDIS code for calculating the ED characteristics for 
several heavy-ion beams. The model is in a reasonable agreement with 
experimental data, when available. We have also made predictions for the
reactions: 160A GeV Pb+Pb (SPS), 100A+100A GeV Au+Au (RHIC) and 2.75A+2.75A TeV 
Pb+Pb (LHC). The inclusive (multiplicity weighted) cross sections for emitting 
nucleons, pions and nuclear fragments in the electromagnetic dissociation of
one of the colliding Au nuclei are shown in Fig. 2 as functions of the incident 
c.m. energy.
Nuclear fragments are divided in two groups: fission fragments (30$<$Z$\le$ 50)
and Intermediate Mass Fragments (IMF´s, 3$\le$Z$\le$30), which are associated
with the multifragmentation. One can clearly see a steep rise in the yields of
all species, especially IMF´s, when the incident energy grows from the SPS to
RHIC and LHC domain. The inclusive cross section for neutron emission is 
especially large, above 1000 b at RHIC and LHC. The average neutron 
multiplicities are predicted to be 4.1, 7.2 and 8.8 at SPS, RHIC and LHC 
respectively. 

The predicted neutron multiplicity distributions are shown in Fig. 3. They 
have a nontrivial structure. There is a strong peak at 1n emission channel
associated with the GR decay. On the other hand, there is a long tail of
multiple neutron emission associated with more violent reaction channels,
from the direct knock-out and evaporation from the compound nucleus to fission
and multifragmentation. This is where our model including all these channels 
shows its strength. One can see, for example, that the probability to emit
more than 20 neutrons is quite noticeable ($\approx 5\%$ at RHIC). These results
might be important for designing neutron-sensitive zero-degree calorimeters
at RHIC and LHC. One of such proposals was made recently in ref. \cite{baltz}
but only the 1n channel was considered there.    
\\

\begin{center}
{\large \bf FIRST ORDER PHASE TRANSITION IN FAST DYNAMICS}
\end{center}

The implications of a strong collective expansion on the liquid-gas phase 
transition were discussed in ref. \cite{nn97}. Here I will focus on 
consequences of the strong collective flow of matter for a possible first 
order chiral transition. I will assume that the collective velocity field is
described locally by the Hubble law, $v(r)=H\cdot r$, where the Hubble
``constant'' $H$ may in general depends on time. 

To make the discussion more concrete, I adopt a picture of the chiral phase 
transition predicted by the linear sigma-model with constituent
quarks \cite{Moc}. Then the mean chiral field $\Phi=(\sigma,{\bf 
\pi})$ serves as an order parameter. The model respects 
chiral symmetry, which is spontaneously broken in the vacuum where 
$\sigma=f_{\pi}$, ${\bf \pi}=0$. 
The effective thermodynamic potential $\Omega(T,\mu;\Phi)$ depends,
besides $\Phi$, on temperature $T$ and baryon chemical potential $\mu$.  
The schematic behaviour of $\Omega(T,\mu;\Phi)$ as a function of the order
parameter field $\sigma$ at $\pi=0$ is shown in Fig. 4. 
The minima of $\Omega$ correspond to the stable or metastable states of 
matter under the condition of thermodynamical equilibrium, where the pressure 
is $P=-\Omega_{min}/V$. 
The curves from bottom to top correspond to different stages of the 
isentropic expansion of homogeneous matter. 
Each curve represents a certain point on the ($T,\mu$) trajectory. 
As one can see from the figure, the model of ref. \cite{Moc} reveals a rather 
weak first order phase transition, although some other models \cite{Raj,Jac} predict a stronger
transition. The discussion below is quite general.
\begin{figure}
\vspace{-2.5cm}
%\begin{center}
%\vspace{-2cm}
\hspace{2cm}
\epsfxsize=10cm
\epsffile{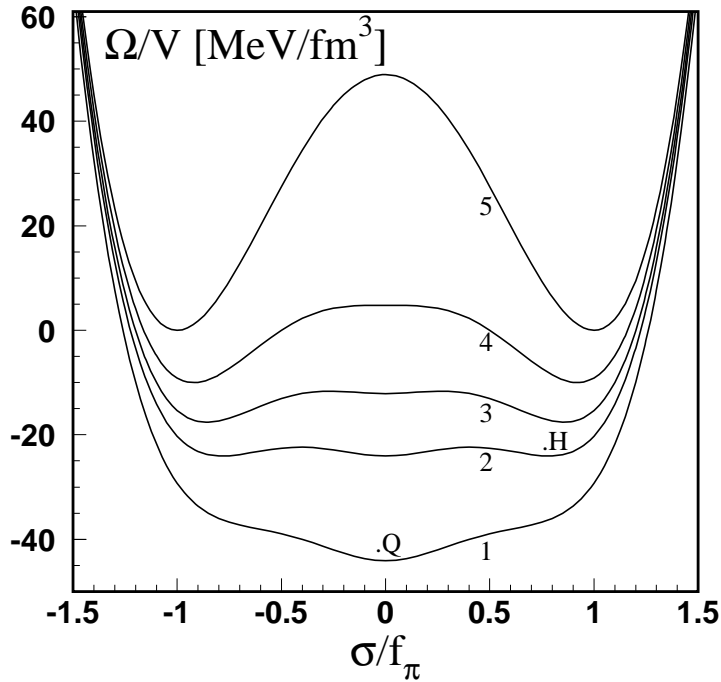}
%\end{center}
\vspace{-3cm}
\caption{\small
Schematic view of the effective thermodynamic potential per volume
$\Omega/V$ as a function of the order parameter field $\sigma$ at
${\bf \pi}=0$, as predicted by the linear $\sigma$-model in the chiral limit 
$m_\pi=0$ [35]. The curves from bottom to top correspond to the 
different stages of the isentropic expansion of homogeneous matter starting 
from $T$=100 MeV and $\mu$=750 MeV (curve 1). The upper 
curve 5 is the vacuum potential. The other curves are discussed in the text.}
%\end{center}
\end{figure}
 
Assume that at some early stage of the reaction the thermal equilibrium is 
established, and partonic matter 
is in a ``high energy density'' phase Q. This state corresponds to the 
absolute minimum of $\Omega$ with the order parameter 
close to zero, $\sigma\approx 0$, ${\bf \pi}\approx 0$, and chiral symmetry 
restored (curve 1). Due to a very high internal pressure, Q matter will 
expand and cool down.
At some stage a metastable minimum appears in $\Omega$ 
at a finite value of $\sigma$ corresponding to a ``low energy density''
phase H, in which chiral symmetry is spontaneously broken. 
At some later time, the critical line in the ($T,\mu$) plane is crossed 
where the Q and H minima have equal depths, i.e. $P_{\rm H}=P_{\rm Q}$ 
(curve 2). At later times the H phase 
becomes more favorable (curve 3), but the two phases are still 
separated by a potential barrier. 
If the expansion of the Q phase continues until the barrier vanishes 
(curve 4), the system will find itself in an absolutely unstable state at 
a maximum of the thermodynamic potential. Therefore, it will  
freely roll down into the lower energy state corresponding to the H phase.
This situation is known as a spinodal instability.  

As well known, a first order phase transition proceeds through the 
nucleation process. According to the standard theory of homogeneous nucleation
\cite{Kap}, supercritical bubbles of the H phase appear only below the critical
line, when $P_H>P_Q$. 
In rapidly expanding matter the nucleation picture might be very different.
As shown in ref. \cite{Mish0}, the phase separation in this case can start as 
early as the metastable H state appears in the thermodynamic potential, and a 
stable interface between the two phases may exist. An appreciable amount of 
nucleation bubbles and even empty cavities may be created already above the 
critical line. 

The bubble formation and growth will also continue below the critical line. 
Previously formed bubbles will now grow faster due to increasing pressure 
difference, $P_{\rm H}-P_{\rm Q}>0$, between the two phases. 
It is most likely  that the conversion of Q matter on the bubble boundary
is not fast enough to saturate the H phase. Therefore, a fast expansion  
may lead to a deeper cooling of the H phase inside the bubbles compared to 
the surrounding Q matter. Strictly speaking, such a system cannot be
characterized by the unique temperature. At some stage the H bubbles 
will percolate, and the topology of the system will change to isolated 
regions of the Q phase (Q droplets) surrounded by the undersaturated 
vapor of the H phase. 

The characteristic droplet size can be estimated by applying the energy 
balance consideration, proposed by Grady \cite{Gra,Hol} in the study of 
dynamical fragmentation of fluids. The idea is that the fragmentation 
of expanding matter is a local process minimizing the sum of surface and 
kinetic (dilational) energies per fragment volume. As shown in ref. \cite{nn97},
this prescription works fairly well also for multifragmentation of expanding 
nuclei, where the standard~statistical~approach~fails. 
 
Let us imagine an expanding spherical Q droplet of radius $R$, embedded in the 
background of the dilute H phase. 
The change of the thermodynamic potential, $\Delta \Omega$, compared to the 
uniform H phase can be easily estimated within the thin-wall approximation 
\cite{Mish0}. According to the Grady's prescription, the quantity to be 
minimized is $\Delta \Omega$ per droplet volume, $V\propto R^3$, that is 
\begin{equation}
\left(\frac{\Delta\Omega}{V}\right)_{droplet}=
-\left(P_{\rm Q}-P_{\rm H}\right)+\frac{3\gamma}{R}
+\frac{3}{10}\Delta {\cal E} H^2 R^2~. 
\end{equation}
Here $\Delta \cal{E}= \cal{E}_{\rm Q}-\cal{E}_{\rm H}$ is the difference of
the bulk energy densities of the two phases, $\gamma$ is the interface energy
per unit area. One should notice that the last term, i.e. the change in the 
collective kinetic energy, is positive because $\cal{E}_{\rm Q}>\cal{E}_{\rm H}$.
This term acts here as an 
effective long-range potential, similar to the Coulomb potential in nuclei.
Since the bulk term does not depend on $R$ the minimization condition 
constitutes the balance between the collective kinetic energy and the interface 
energy. This leads to an optimum droplet radius
\begin{equation}  \label{R}
R^{\ast}=\left(\frac{5\gamma}{\Delta {\cal E} H^2}\right)^{1/3}.
\end{equation}
One can say that the metastable Q phase is torn apart by a 
mechanical strain associated with the collective expansion. 
This phenomenon has a direct analogy with the fragmentation of pressurized 
fluids leaving nozzles \cite{Bli,Toe}. In a similar way, splashed water forms
droplets which have little to do with the equilibrium liquid-gas phase 
transition. 

In the lowest-order approximation the characteristic droplet mass can be
calculated as  $M^\ast\approx\Delta{\cal E}V$. It is natural to think that
nucleons and heavy mesons are smallest droplets of the Q phase. 
For numerical estimates I  take $\gamma=10$ MeV/fm$^2$ and $\Delta{\cal E}=0.5$ 
GeV/fm$^3$, i.e. the energy density inside the nucleon. For the Hubble constant 
I consider two possibilities: $H^{-1}=20$ fm/, representing a slow expansion 
from a soft point, and $H^{-1}=6$ fm/c typical for a fast expansion.
Substituting these values in Eq. (\ref{R}) one gets $R^\ast$=3.4 fm and 1.5 fm 
for the slow and fast expansion respectively. These two values of $R^*$ give
$M^\ast$ of about 100 GeV and 10 GeV, respectively. 
At ultrarelativistic energies the collective expansion is
very anisotropic, with the strongest component along the beam axes.
For the predominantly 1-d expansion one should expect the formation of 
slab-like structures with intermittent layers of Q and H phases.
%Using the minimum information 
%principle one can show \cite{Hol,nn97} that the distribution of droplets 
%should follow an exponential law, $\exp{\left(-{M \over M^{\ast}}\right)}$. 
%Thus, about 2/3 of droplets have masses smaller than $M^*$, but 
%Therefore, with 1$\%$ probability one
%can  find droplets as heavy as $5M^{\ast}$. 

After separation the droplets recede from each other according to the 
global Hubble expansion, pre\-do\-mi\-nant\-ly along the beam direction. 
Hence their center-of-mass 
rapidities are in one-to-one correspondence with their spatial positions.
Presumably they will be distributed more or less evenly between 
the target and projectile rapidities. 
Since rescatterings in the dilute H phase are rare, most hadrons produced 
from individual droplets will go directly into detectors.  One can guess that
the number of produced hadrons is proportional to the droplet mass. Each 
droplet will give a bump in the hadron rapidity distribution around its 
center-of-mass rapidity. If emitted particles have a Boltzmann spectrum, the 
width of the bump will be $\delta y \sim 2\sqrt{T/m}$, where $T$ is the 
droplet temperature and $m$ is the particle mass. At $T\sim 100$ MeV this 
gives $\delta y\approx 2$ for pions and $\delta y\approx 1$ for nucleons. 
These spectra might be slightly modified by the residual expansion of 
droplets and their transverse motion. The resulting rapidity distribution 
in a single event will be a superposition of 
contributions from different droplets, and therefore it will exhibit strong 
non-statistical fluctuations. The fluctuations will be more pronounced 
if primordial droplets are big, as expected in the vicinity of the soft point.
If droplets as heavy as 100 GeV are formed, each of them will emit up to 
$\sim$300 pions within a narrow rapidity interval, $\delta y\sim 1$. 
Such bumps can be easily resolved and analyzed. The fluctuations will be less
pronounced if many small droplets shine in the same rapidity interval. 
Critical fluctuations of similar nature were discussed in ref. \cite{Ant}.
  
Some unusual events produced by high-energy cosmic nuclei have been already 
seen by the JACEE collaboration \cite{JACEE}. Unfortunately, they are very few 
and it is difficult to draw definite conclusions by analyzing them. 
We should be prepared to see plenty of such events in the future RHIC and 
LHC experiments.
It is clear that the nontrivial structure of the hadronic spectra will be 
washed out to a great extent when averaging over many events. Therefore, more 
sophisticated methods of the event sample analysis should be used. 
The simplest one is to  search for non-statistical fluctuations in the 
hadron multiplicity distributions measured in a fixed rapidity bin
\cite{Tan}. One can also study the correlation of multiplicities in 
neighbouring rapidity bins, bump-bump correlations etc. 
Such standard methods as intermittency and commulant moments \cite{Ant}, 
wavelet transforms \cite{Suz}, HBT interferometry \cite{Hei} can also be
useful. All these studies should be done at 
different collision energies to identify the phase transition threshold.
The predicted dependence on the Hubble constant and the reaction geometry 
can be checked in collisions with different ion masses and 
impact parameters.
\\

\begin{center}
{\large \bf CONCLUSIONS}
\end{center}
%\\
%The main conclusions of the talk are as follows:
\begin{itemize}

\item The statistical approach (SMM) works well in situations when thermalized 
sources are well defined and no significant collective flow is present.

\item The quantitative agreement of SMM with recent data on the caloric curve
and temperature fluctuations provides a strong indication on the nuclear
liquid-gas phase transition. The nuclear heat capacity has a peak at $T\approx$
6 MeV.

\item A first order phase transition in rapidly expanding matter should proceed
through the nonequilibrium stage when a metastable phase splits into droplets. 
The primordial droplets should be biggest in the vicinity of a soft point when
the expansion is slowest.

\item Hadron emission from droplets of the quark-gluon plasma should lead to
large nonstatistical fluctuations in their rapidity spectra and multiplicity
distributions. The hadron abundances may reflect directly the chemical
composition in the plasma phase.

\item Electromagnetic excitation of nuclei in ultrarelativistic heavy-ion
colliders is an important reaction mechanism leading to the deep nuclear
disintegration. The multiple neutron emission associated with this process may
be used for monitoring ultrarelativistic heavy-ion beams.

\item And finally, we should use the lessons of the liquid-gas phase transition 
for future studies of the deconfinement-hadronization and chiral phase 
transitions in relativistic heavy-ion collisions.
   
\end{itemize}
\vspace{0.3cm}
\begin{center}
{\large \bf ACKNOWLEDGMENTS}
%\vspace{0.5 truecm}
\end{center}
\nopagebreak
The author is grateful to J.P. Bondorf and A.D. Jackson for many fruitful 
discussions. I thank A.S. Botvina, M. D'Agostino, A. Mocsy, I.A. Pshenichnov, 
O. Scavenius  for cooperation. Discussions with D. Diakonov, A. Dumitru, 
J.J. Gaardh{o}je, M.I. Gorenstein, W. Greiner, B. Jakobsson, L. McLerran, 
R. Mattiello, W.F.J. M\"uller, W. Reisdorf, L.M. Satarov, H. St\"ocker, 
E.V. Shuryak, W. Trautmann and V. Viola are greatly appreciated. I thank the 
Niels Bohr Institute, Copenhagen University, and the Institute for Theoretical 
Physics, Frankfurt University, for kind hospitality. This work was carried out 
partly within the framework of a Humboldt Award, Germany.
%\begin{center}
%{\Large \bf References}
%\end{center}
\vspace{-0.3cm}


\begin{thebibliography}{99}

{\small \baselineskip=10pt
\bibitem{bondorf3} J.P. Bondorf, A.S. Botvina and I.N. Mishustin, {\it Phys.
Rev.} {\bf C58}, R27 (1998).
\bibitem{Dago99} M. D'Agostino, A.S. Botvina, M. Bruno, A. Bonasera, 
J.P. Bondorf, I.N. Mishustin, F. Gulminelli, R. Bougault, N. Le Neindre, 
P, Desesquelles, E. Geraci, A. Pagano, I. Iori, A. Moroni, G.V. Margagliotti,
G. Vannini, {\it Nucl. Phys.} {\bf A650}, 329 (1999).
\bibitem{I} J.P. Bondorf, R. Donangelo, I.N. Mishustin, C.J. Pethick, H. Schulz,
K. Sneppen, {\it Nucl. Phys.} {\bf A443}, 321 (1985).
\bibitem{Mish85} I.N. Mishustin, {\it Nucl. Phys.} {\bf A 447} (1985) 67c.
\bibitem{Raj} J. Berges and K. Rajagopal, {\it Nucl. Phys.} {\bf B538}, 215
(1999).
\bibitem{Jac} M.A. Halasz, A.D. Jackson, R.E. Shrock, M.A. Stephanov and 
J.J.M. Verbarshot, {\it Phys. Rev.} {\bf D58}, 096007 (1998).
%\bibitem{Car} G. Carter and D. Diakonov, {\it Nucl. Phys.} {\bf A} (in press); 
%hep-ph/9807219. 
\bibitem{Ste} M. Stephanov, K. Rajagopal and E. Shuryak, {\it Phys. Rev. Lett.} 
{\bf 81}, 4816 (1998).
\bibitem{Sat} I.N. Mishustin, L.M. Satarov, H. Stoecker and W. Greiner, {\it
Phys. Rev.} {\it C59}, 3243 (1999).
\bibitem{reisdorf} W. Reisdorf and FOPI Collaboration, {\it Nucl. Phys.} {\bf A612},
493 (1997).
\bibitem{Bra} P. Braun-M\"unzinger and J. Stachel, 
%in Proceedings of the 13th 
%International Conference on Ultrarelativistic Nucleus-Nucleus Collisions, 
%``Quark Matter 97'' (Tsukuba, December 1-5, 1997); 
{\it Nucl. Phys.} {\bf A638}, 3c (1998).
\bibitem{Mish0} I.N. Mishustin, {\it Phys. Rev. Lett.} {\bf 82}, 4779 (1999).
\bibitem{Pshe1} I.A. Pshenichnov, I.N. Mishustin, J.P. Bondorf, A.S. Botvina,
A.S. Iljinov, {\it Phys. Rev.} {\bf C57}, 1920 (1998).
\bibitem{Pshe2} I.A. Pshenichnov, I.N. Mishustin, J.P. Bondorf, A.S. Botvina,
A.S. Iljinov, {\it Phys. Rev.} {\bf C60}, 044901 (1999).
\bibitem{PR} J.P. Bondorf, A.S. Botvina, A.S. Iljinov, I.N. Mishustin and 
and K.Sneppen, {\it Phys. Rep.} {\bf 257} (1995) 133.
\bibitem{gross} D.H.E. Gross, {\it Rep. Progr. Phys.} {\bf 53} (1990) 605.
\bibitem{das} S. Das Gupta, J. Pan, I. Kvasnikova, C. Gale, {\it Nucl. Phys.} 
{\bf A621}, 897 (1997). 
\bibitem{botvina1} A.S. Botvina et al., {\it Nucl. Phys.} {\bf A584} (1995) 737.
\bibitem{xi}H. Xi and ALADIN Collaboration, {\it Z. Phys.} {\bf A359}, 397 (1997).
\bibitem{dagostino} M. D'Agostino et al., {\it Phys. Lett.} {\bf B371},175 (1996).
\bibitem{boug}R. Bougault et al., {\it in Proceedings of the XXXV International
Winter Meeting on Nuclear Physics (Bormio, February 3-7, 1997)}; {\it Preprint} 
LPCC 97-04, April 1997.
\bibitem{viola} V.E. Viola et al., {\it in Proceedings of the International Workshop
on Gross Properties of Nuclei and Nuclear Excitations XXVII: Multifragmentation
(Hirschegg, Austria, 17-23 January 1999)}, p. 93; {\it Preprint} INC-40007-136, Indiana
University, 1999. 
\bibitem{sri} R.P. Scharenberg, B.K. Srivastava and EOS Colaboration, {\it in 
the same Hirschegg Proceedings as above}, pp. 237, 247.
\bibitem{bondorf} J.P. Bondorf, R. Donangelo, I.N. Mishustin, H. Schulz, 
{\it Nucl. Phys.} {\bf A444}, 460 (1985).
\bibitem{pochod} J. Pochodzalla and ALADIN Collaboration, {\it Phys. Rev. Lett.} {\bf 75}, 1040 (1995).
\bibitem{mul} W.F.J. M\"uller, {in the same Hirschegg Proceedings as above}, p.
200.
\bibitem{albergo} S. Albergo et al., {\it Nuovo Cimento} {\bf 89}, 1 (1985).
\bibitem{tsang} M.B. Tsang, W.G. Lynch, H. Xi and W.A. Friedman, 
{\it Phys. Rev. Lett.} {\bf 78} 3836 (1997).
\bibitem{raduta} Al. H. Raduta, Ad.R. Raduta, {\it Phys. Rev.} {\bf C59}, 323 
(1999).
\bibitem{campi} X. Campi, {\it J. Phys.} {\bf A19}, L917 (1986); {\it Phys.
Lett.} {\bf B208}, 351 (1988).
\bibitem{fisher} M.E. Fisher, {\it Rep. Prog. Phys.} {\bf 30}, 615 (1967).
\bibitem{chomaz} F. Gulminelli and Ph. Chomaz, {\it Phys. Rev. Lett.} {\bf 82},
1402 (1999).
\bibitem{Pshe0} A.S. Iljinov, I.A. Pshenichnov, N. Bianchi, E. De Sanctis, V.
Muccifora, M. Mirazita and P. Rossi, {\it Nucl. Phys.} {\bf A616}, 575 (1997).
\bibitem{baltz} A.J. Baltz, C. Chasman,and S.N. White, nucl-ex/9801002.
\bibitem{nn97} I.N. Mishustin, 
{\it in Proceedings of the 6th International Conference
on Nucleus-Nucleus Collisions (Gatlinburg, June 2-6, 1997)}; 
{\it Nucl. Phys.} {\bf A630}, 111c (1998).
\bibitem{Moc} L.P. Csernai, I.N. Mishustin and A. Mocsy, {\it Heavy Ion Phys.},
{\bf 3}, 151 (1996);\\ A. Mocsy, M.Sc. thesis, University of Bergen, 1996. 
\bibitem{Kap} L.P. Csernai, J.I. Kapusta, {\it Phys. Rev. Lett.} {\bf 69}, 737 
(1992); {\it Phys. Rev.} {\bf D46}, 1379 (1992).
\bibitem{Gra} D.E. Grady, {\it J. Appl. Phys.} {\bf 53}(1), 322 (1981).
\bibitem{Hol} B.L. Holian and D.E. Grady, {\it Phys. Rev. Lett.} {\bf 60}, 1355 
(1988).
\bibitem{Bli} J.A. Blink and W.G. Hoover, {\it Phys. Rev.} {\bf A32}, 1027 (1985).
\bibitem{Toe} H.Buchenau et al., 
%E.L. Knuth, J. Northby J.P. Toennies, C. Winkler,
{\it J. Chem. Phys.} {\bf 92}, 6875 (1990). 
\bibitem{Ant} N.G. Antoniou, {\it Nucl. Phys.} {\bf B71}, 307 (1999).
\bibitem{JACEE} T.H. Barnett et al., {\it Phys. Rev. Lett.} {\bf 50}, 2062 (1983).
\bibitem{Tan} M.J. Tannenbaum and E802 Collaboration, {\it Phys. Rev.} {\bf C52}, 
2663 (1995)
\bibitem{Suz} N. Suzuki, M. Biyajima and A. Ohsawa, hep-ph/9503403. 
\bibitem{Hei} H. Heiselberg, A.D. Jackson, hep-ph/9809013.
}

\end{thebibliography}
\end{document}